\newcommand{\PHCC} {(C$_4$H$_{12}$N$_2$)Cu$_2$Cl$_6$ \xspace}

\newcommand{\TCL} {TlCuCl$_3$\xspace}
\newcommand\musr{$\mu$SR\xspace}

\documentclass[prl,twocolumn,floatfix,preprintnumbers,amsmath,amssymb,superscriptaddress]{revtex4}

\usepackage{graphicx}
\usepackage{dcolumn}
\usepackage{bm}
\usepackage{color}
\usepackage{xspace}

\begin{document}

\title{Pressure-Induced Quantum Critical and Multicritical Points in a Frustrated Spin Liquid}

\author{M. Thede}
\affiliation{Neutron Scattering and Magnetism, Laboratory for Solid State
Physics, ETH Z\"urich, Z\"urich, Switzerland}
\affiliation{Laboratory for Muon Spin Spectroscopy, Paul Scherrer Insitut,
Villigen-PSI, Switzerland}

\author{A. Mannig}
\affiliation{Neutron Scattering and Magnetism, Laboratory for Solid State
Physics, ETH Z\"urich, Z\"urich, Switzerland}

\author{M. M{\aa}nsson}
\altaffiliation[Present address: ]{Laboratory for Quantum Magnetism (LQM)
\'Ecole polytechnique f\'ed\'erale de Lausanne (EPFL), Lausanne, Switzerland and Laboratory for Neutron Scattering, Paul Scherrer Insitut, Villigen, PSI, Switzerland}

\affiliation{Neutron Scattering and Magnetism, Laboratory for Solid State
Physics, ETH Z\"urich, Z\"urich, Switzerland}

\author{D. H\"uvonen}
\affiliation{Neutron Scattering and Magnetism, Laboratory for Solid State
Physics, ETH Z\"urich, Z\"urich, Switzerland}

\author{R. Khasanov}
\affiliation{Laboratory for Muon Spin Spectroscopy, Paul Scherrer Insitut,
Villigen-PSI, Switzerland}

\author{E. Morenzoni}
\affiliation{Laboratory for Muon Spin Spectroscopy, Paul Scherrer Insitut,
Villigen-PSI, Switzerland}

\author{A. Zheludev}
 \email{zhelud@ethz.ch}
 \homepage{http://www.neutron.ethz.ch/}
\affiliation{Neutron Scattering and Magnetism, Laboratory for Solid State
Physics, ETH Z\"urich, Z\"urich, Switzerland}

\date{\today}

\begin{abstract}
The quantum spin-liquid compound \PHCC is studied by \musr under hydrostatic pressures up to 23.6~kbar. At low temperatures, pressure-induced incommensurate magnetic order is detected beyond a quantum critical point at $P_c\sim4.3$~kbar. An additional phase transition to a different ordered phase is observed at $P_1\sim 13.4$~kbar. The data indicate that the high-pressure phase  may be a commensurate one. The established $(P-T)$ phase diagram reveals the corresponding pressure-induced multicritical  point at $P_1$, $T_1=2.0$~K.
\end{abstract}

 \maketitle

Traditionally, magnetic insulators have been the most important prototype systems for testing concepts and theories of phase transitions, universality and scaling \cite{CollinsBook,StanleyBook}. They owe this to their well-defined short-range interactions, a broad range of interaction topologies and dimensionalities, and their amenability to numerical modeling. With a more recent interest in quantum phase transitions \cite{Sachdevbook,Sachdev2011}, magnetic insulators have become the prototypes of choice to study quantum critical points (QCPs). Realizations of such  important QCPs as Bose-Einstein condensation (BEC) \cite{Giamarchi2008}, deconfinement in one dimension \cite{Thielemann2009,Schmidiger2013-2}, and the Ising model in a transverse field  \cite{Coldea2010} have been found in quantum magnets in applied magnetic fields. Magnetic BEC, for example, occurs in gapped quantum  antiferromagnets (AFMs) with a spin singlet ground state, when an external field drives the energy gap to zero by virtue of Zeeman effect. The result in spontaneous long-range magnetic order in the perpendicular direction, and thus a breaking of SO(2) symmetry \cite{Giamarchi2008}. At the QCP, the soft mode has a parabolic dispersion, so that the dynamical critical exponent is $z=2$. By now, this transition has been extensively studied experimentally and theoretically \cite{Giamarchi2008}.

A qualitatively different type of soft mode transition in gapped quantum AFMs may occur if the spin gap is driven to zero by varying the ratio of exchange constants. The resulting spontaneous long-range magnetic order breaks $SO(3)$ symmetry, and the spectrum is expected to be linear at the QCP ($z=1$).  In practice, the only way to continuously tweak the exchange interactions is by applying external pressure. Closing the spin gap with pressure in quantum Heisenberg AFMs has been  attempted in experiments \cite{Zaliznyak1998,Hong2008,Hong2010}. However, only one good  realization of pressure-induced ordering in such systems has been found to date, namely, that in \TCL \cite{Tanaka2003,oosawa2004,Ruegg2004}. Further studies of this QCP brought fascinating new insights \cite{Ruegg2008}, particularly the observation of a longitudinal mode, which is a magnetic analog of the celebrated Higgs boson \cite{Sachdev2011}. In the present work, we report the observation of pressure-induced ordering in the $S=1/2$ frustrated gapped quantum AFM \PHCC (abbreviated PHCC), and use muon spin rotation (\musr) experiments to map out the $P-T$ phase diagram. We show that the pressure-driven transition leads to an {\it incommensurate} magnetic order. At still higher pressures, we detect an {\it additional} transition and  multicritical point. The indication is that these are an incommensurate to commensurate (IC) transformation, and the associated Lifshitz point.
\begin{figure}
\includegraphics[width=\columnwidth]{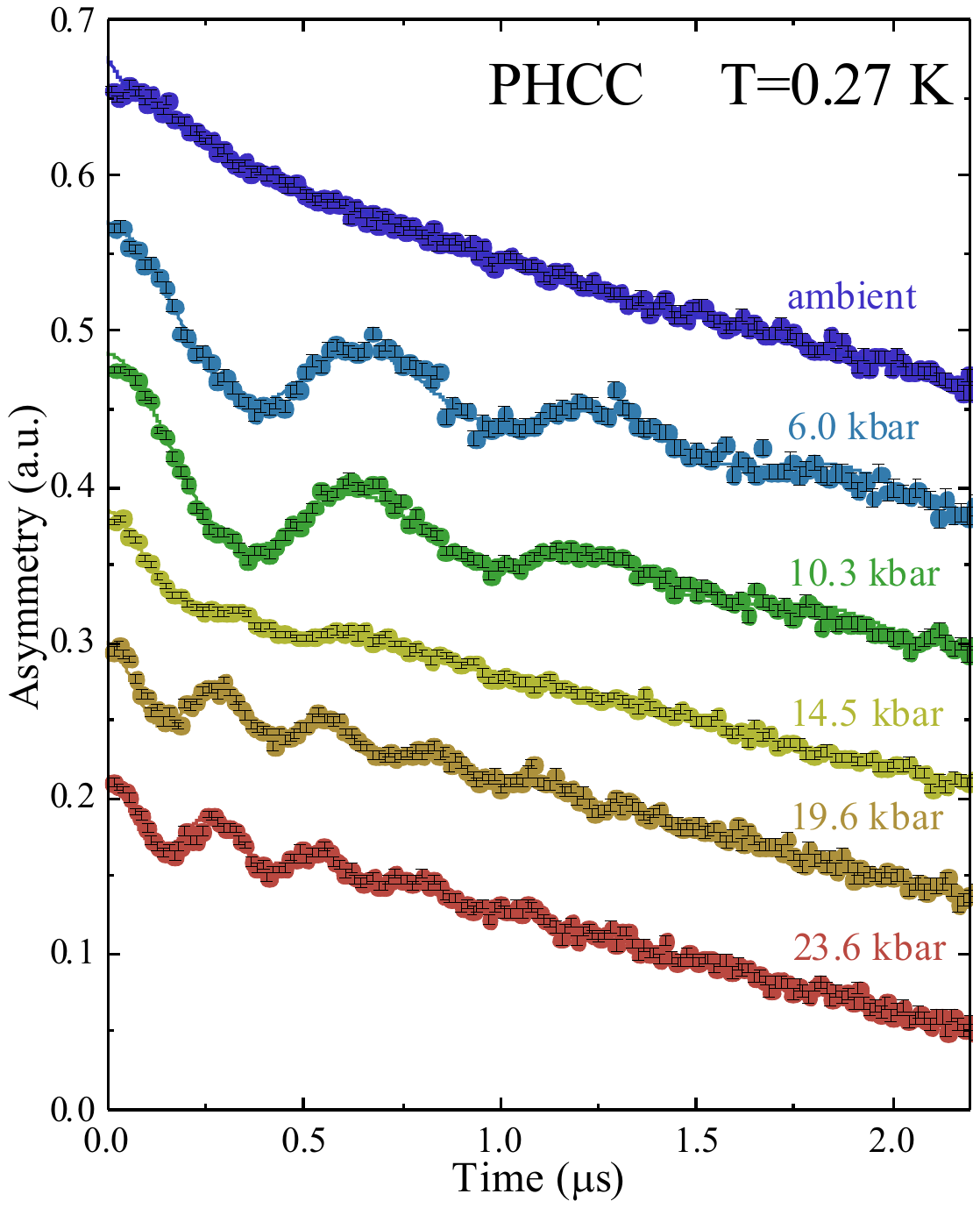}
\caption{(Color online)  Symbols: typical \musr spectra measured in PHCC at $T=0.27$~K under different applied hydrostatic pressures.  Lines are fits to the data as described in the text. From the bottom up, the data are
offset along the y axis by -0.09, 0, +0.09, 0.18, 0.27, and 0.36 \label{Fig:spectra}}
\end{figure}

The magnetic and spectroscopic properties of PHCC have been studied extensively. The crystal structure is triclinic, where S = 1/2-carrying Cu$^{2+}$ ions are connected into a rather complex layered
spin network, with a high degree of geometric frustration \cite{Stone2001}. The ground state is a non-magnetic spin singlet with only short-range correlations, known as a quantum ``spin liquid'' \cite{Stone2001,Stone2006}. The lowest energy excitations are a $S=1$ triplet, with a gap $\Delta=1.0$~meV and a rather narrow bandwidth of 1.7~meV \cite{Stone2001,Stone2006-Nature}. The dispersion is a global minimum at the AF zone-center $(1/2,0,1/2)$. A BEC-type magnetic ordering with this wave vector can be induced in PHCC by the application of a magnetic field exceeding $H_c=7.5$~T \cite{Stone2006,Stone2007,Huevonen2013}. Previously, PHCC was investigated under hydrostatic pressure using inelastic neutron scattering \cite{Hong2010}. The gap was found to decrease with increasing pressure. A linear extrapolation of this dependence, which was measured up to 10~kbar, suggested a gap closure and QCP at around 20~kbar. It is that transition that we set out to look for with \musr.

\musr is a technique that is exceptionally sensitive to very small magnetic moments, and is therefore a useful tool in the study of quantum magnets \cite{Lancaster2006,Sugano2010,Pratt2006}. In addition, the method can be applied in bulky sample environments such as pressure cells \cite{Bendele2010,Ghannadzadeh2013}. In our experiments, we used powdered solution-grown PHCC polycrystalline samples of typical mass 800~mg. The measurements were performed in MP35 and CuBe piston-cylinder clamp cells,  specifically designed for \musr experiments \cite{Andreica2001}.
The pressure medium was Daphne Oil 7373, and low temperatures were achieved using a $^3$He cryostat. The actual sample pressure was determined by measuring the $T_c$ shift of an In superconducting sensor by means of AC-susceptibility. The data were collected on the GPD instrument on the $\mu$E1 beamline at the S$\mu$S muon source at Paul Scherrer Institut. The analysis was performed using \textsc{Musrfit} \cite{Suter2012}.

Spin-polarized muons produced by the accelerator facility are implanted in the sample. One measures the real-time Larmor precession of the muon spins in the local magnetic fields at their resting sites. In the experiment, depending on the type of pressure cell used, 50--70\% of the muons stop in the thick walls of the pressure cell, leading to a background signal, whose functional form is known for each cell type, so that it can be easily accounted for in the data analysis. In the case of a magnetically ordered sample, the spins muons stopped in the sample will precess coherently in the local field.  Their polarization will show an oscillating component with a frequency proportional to the local field at the stopping site. The latter is proportional to the size of the spontaneously ordered moment, so the oscillation frequency can be taken as a measure of the order parameter.\cite{YaouancReotier}

Typical precession curves observed at the lowest experimental temperature of 270~mK in PHCC for different applied pressures are shown in Fig.~\ref{Fig:spectra}. Several regimes are apparent. Below 4.4~kbar no oscillations are found. The signal can be well understood as the sum of the pressure cell component and of the sample component whose relaxation is due to nuclear moments and dynamic electron moments. The dynamic character is confirmed by decoupling measurements in longitudinal field showing the absence of static magnetism. Surprisingly, already at $P_c\sim 4.4$~kbar, well below the transition pressure suggested by neutron experiments, clear oscillations are observed. This is an unambiguous sign of long-range magnetic order in the sample. At $P_1\approx 13.4$~kbar the oscillations largely {\it disappear}.The fast initial  polarization decay can not be attributed to nuclear spins alone. This type of behavior indicates a disordered magnetism or is the result of the microscopic co-existence of several distinct magnetic phases. Finally, clear oscillations re-emerge above $P_1$. In this high-pressure regime they are visibly higher than for $P_c<P<P_1$, and, as will be explained below, have a distinct functional form. We conclude that we are dealing with at least three phases: a spin liquid for $P<P_c$, an ordered state ``Phase A'' for $P_c<P<P_1$, a {\it different} ordered ``Phase B'' for $P>P_1$, and perhaps a phase co-existence in the vicinity of $P_1$. The latter can be attributed to the unavoidable pressure gradient across the cell, estimated to be around 0.5~kbar.

\begin{figure}
\includegraphics[width=\columnwidth]{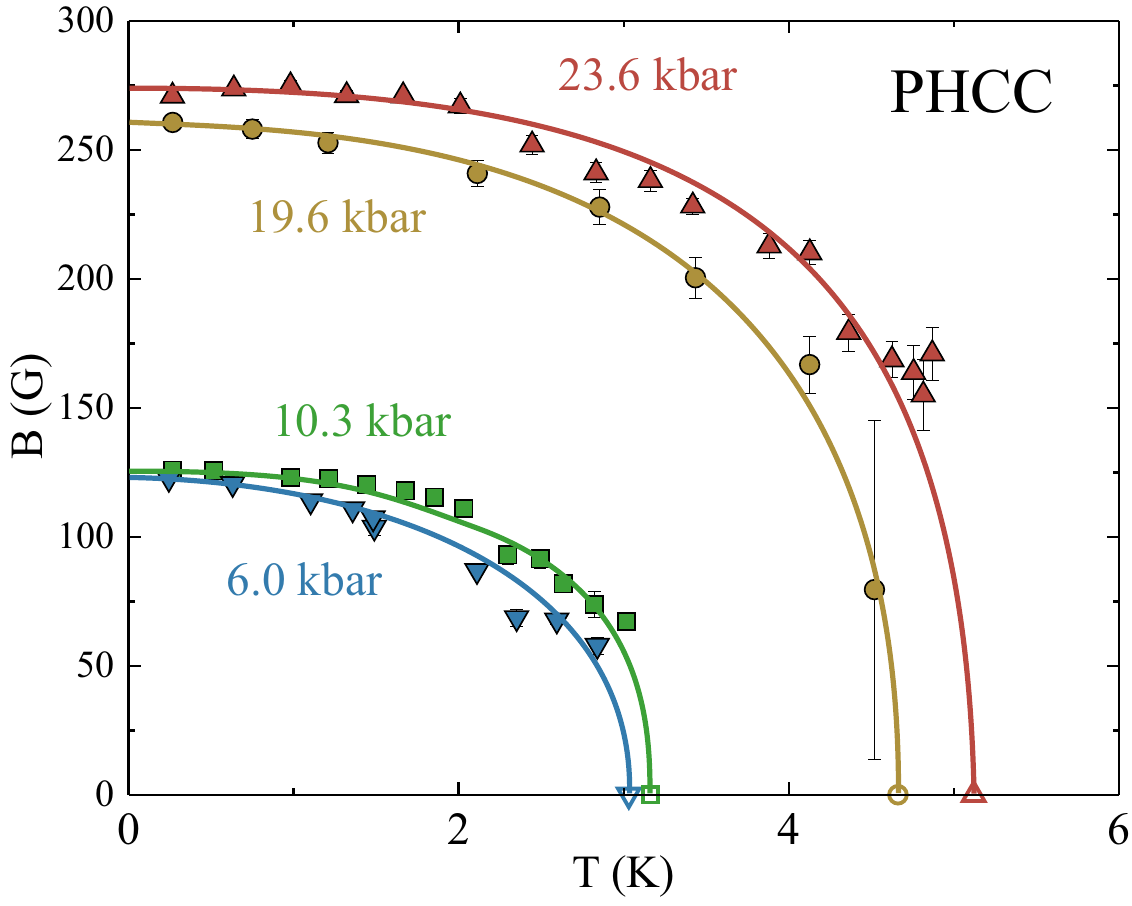}
\caption{(Color online)  Measure temperature dependence of the local Larmor field $B$ at the muon site in PHCC for several representative pressures (solid symbols). Open symbols are transition temperatures, as determined in transverse-field measurements. Solid lines are guides to the eye.\label{Fig:orderp}}
\end{figure}

The nature of the two ordered phases can be elucidated from the functional form of the oscillations. In the simplest scenario of a commensurate structure and all muons resting at magnetically  equivalent sites, the muon time-spectra are expected to be a simple damped cosine-function \cite{YaouancReotier}. The complete functional form used in our data analysis is described in the Supplemental Material. It includes the contribution of the pressure cell, a slow background due to muons on non-magnetic sites, and accounts for a possible narrow inhomogeneity of pressures inside the cell. The main term describing muon precessions in the ordered state, relevant to the present discussion, is \cite{YaouancReotier}:
\begin{equation}
 A(t)=A\left[\frac{2}{3}\cos(\omega t +\phi)\exp(-\lambda_1 t)+\frac{1}{3}\exp(-\lambda_2t)\right].\label{Eq:cos}
\end{equation}
 Here $\omega$ is related to the local field through $\omega=\gamma B$, with $\gamma=85.16\cdot 10^3$ radian s$^{-1}$ G$^{-1}$. The oscillatory term corresponds to the  component of the muon  spin perpendicular to the local static field. The second term represents the loss of polarization due to dynamical effects, for the spin component parallel to the local static field.  For the high-pressure Phase B, Eq.~\ref{Eq:cos} gives good fits to the data with $\phi=0$, as shown in solid lines in Fig.~\ref{Fig:spectra}(19.6~kbar, 23.6~kbar). This strongly suggests that this phase is a simple collinear structure \footnote{ It has to be noted that equally good fits can be obtained by allowing a small phase shift $\phi \sim 20^\circ$, depending on pressure, with essentially the same value of $\chi^2$. Therefore, the incommensurability of Phase B can not be entirely ruled out.}.

Interestingly, the collinear model can {\it not} produce acceptable fits to the oscillations observed in Phase A, unless one introduces a large unphysical phase factor $\phi \sim \text{45}^\circ$ for the cosine. Instead, we found that these spectra can be well fitted if the magnetic part is described by a damped Bessel function:
\begin{equation}
 A(t)=A\left[\frac{2}{3}J_0(\omega t+\phi)\exp(-\lambda_1 t)+\frac{1}{3}\exp(-\lambda_2t)\right].\label{Eq:Bes}
\end{equation}
The corresponding fits are shown in solid lines in Fig.~\ref{Fig:spectra}(6~kbar, 10.3~kbar).
These Bessel-type oscillations strongly suggest an incommensurate magnetic structure such as spin density wave or helimagnet \cite{YaouancReotier}. Incommensurate order produces a continuous distribution of local fields across different crystallographically equivalent muon sites. The Bessel function is a result of beats between frequencies in this continuum, in the simplest case of a sinusoidal spatial modulation \footnote{Unlike for Phase B, the incommensurate nature of phase A is an unambiguous conclusion of the data analysis. It is also clear that Phase B, even if incommensurate, has a very different structure, as compared to Phase A. }. For an insulating and strongly geometrically frustrated magnet such as PHCC, a helimagnetic structure is the most natural candidate for this behavior.

\begin{figure}
\includegraphics[width=\columnwidth]{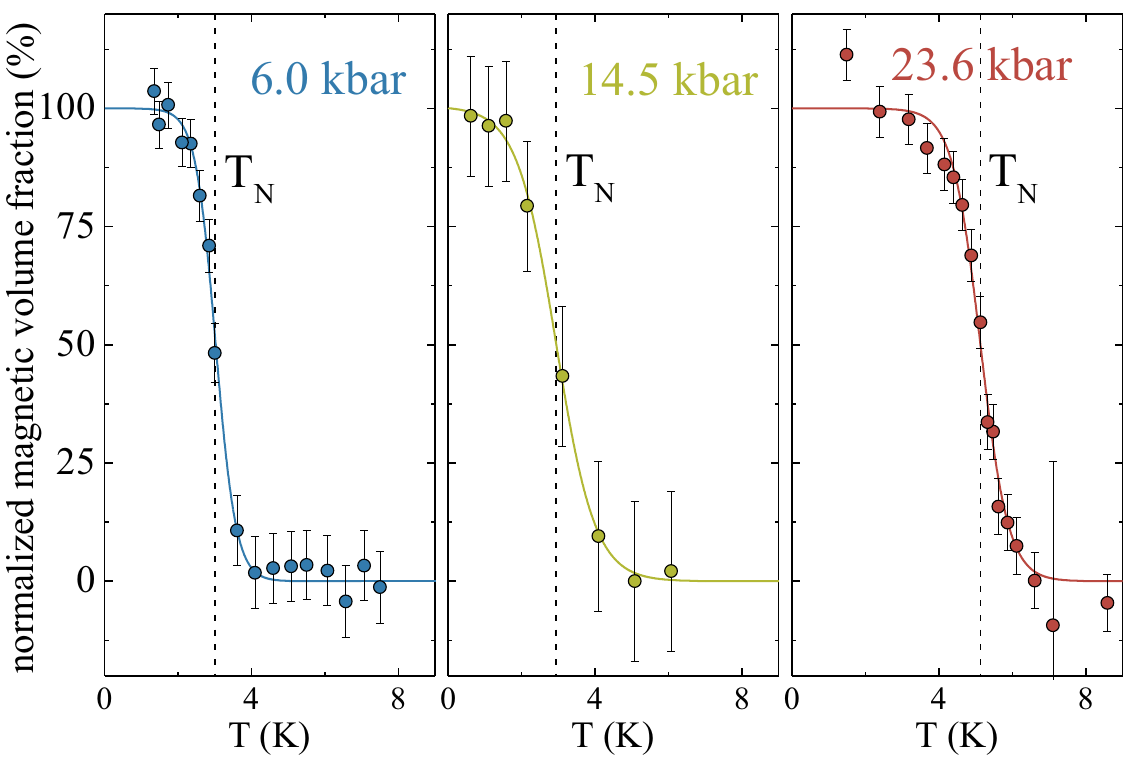}
\caption{(Color online) Typical measured temperature dependence of the normalized magnetic fraction in PHCC at three applied pressures , assuming a 100~\% ordered sample at low temperature. Solid lines are simple sigmoidal fits to the data used to determine the ordering temperatures $T_\mathrm{N}$ (dashed lines).\label{Fig:depol}}
\end{figure}

In each of the two phases, the Larmor frequency $\omega$ can be taken as being proportional to the magnetic order parameter. Its temperature evolution for several applied pressures is shown in Fig.~\ref{Fig:orderp} (symbols). The data analysis described above becomes infeasible close to to the transition temperatures due to the decrease of the local fields. Elsewhere, the temperature dependence of the ordered moment appears to be quite conventional. The saturation moment at low temperatures does not vary much within each phase. However, the typical precession frequencies in Phase B are consistently at least twice as high as in Phase A.

\begin{figure}
\includegraphics[width=\columnwidth]{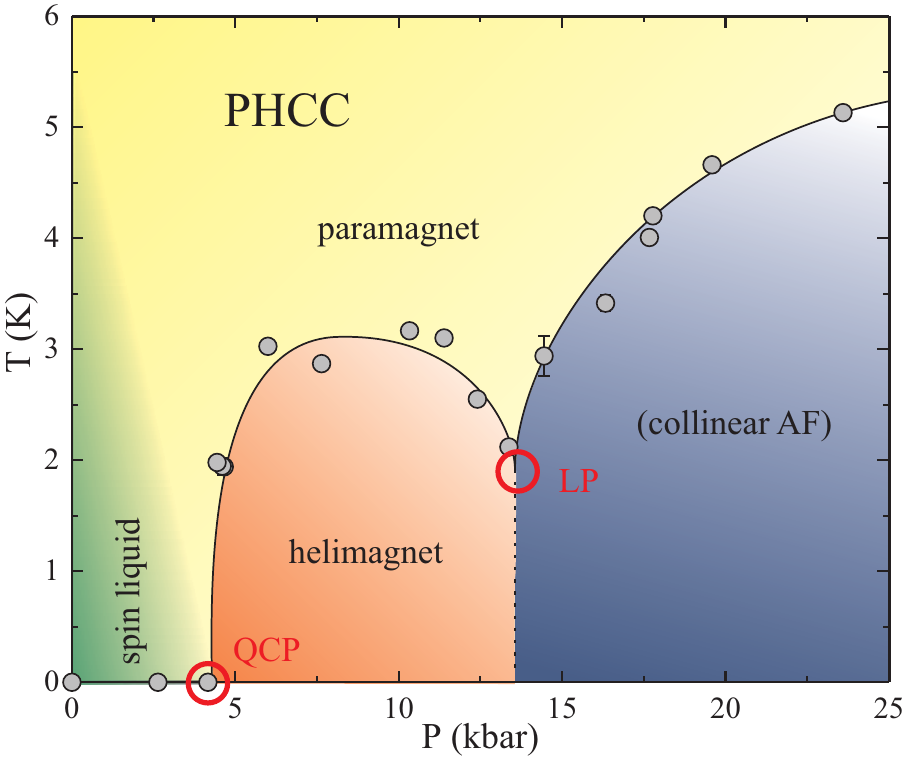}
\caption{(Color online) Measured $P-T$ phase diagram of PHCC. Symbols are transition temperatures, as determined by transverse-field measurements. Symbols at $T=0$ indicate an absence of muon depolarization due to magnetic order. \label{Fig:phase}}
\end{figure}

In order to determine the ordering temperatures more accurately and to map out the $P-T$ phase diagram, we performed experiments in a weak transverse field (wTF)\cite{YaouancReotier}. For this, a small external field of 30 or 50~G is applied to the sample perpendicular to the initial polarization. This measurement allows to determine the magnetic fraction as a function of temperature. When the applied field is larger than the internal static fields sensed by the muons, the amplitude of the asymmetry component oscillating in the applied field represents the non-magnetic volume. In the magnetic phase the internal fields will dominate and dephase the muon ensemble. The magnetic fraction can thus be determined. As shown in Fig.~\ref{Fig:depol}, the onset of magnetic order is marked by a step-like increase of the normalized magnetic volume fraction \footnote{Due to most muons stopping in the pressure cell walls, the actual non-normalized magnetic volume fraction varied between 30\% and 40\% depending on the cell used.}. A similar feature is visible in the vicinity of $P_1$, where the oscillations are not directly observable. As expected, no depolarization is detected in the spin liquid phase. These data were analyzed using an empirical sigmoidal function, which associates the ordering temperature with the center of the drop. Fig.~\ref{Fig:phase} shows the pressure dependence of the ordering temperature (symbols) presented as a phase diagram in the $(P,T)$ plane.

In this diagram it is apparent that the $P_c\sim 4.3$~kbar corresponds to a QCP. We suggest that this ordering transition is due to the closure of the spin gap, caused by a pressure-induced change in the exchange constants. In this case, it is analogous to the one in \TCL in being a $z=1$ QCP with $SO(3)$ symmetry. As mentioned, the most probable cause of incommensurability in the ordered state is geometrically frustrated exchange interactions. These are responsible for helical structures in numerous spin systems, the oldest known example perhaps being MnO$_2$ \cite{Yoshimori1959}. The dispersion of magnetic excitations in PHCC at ambient pressure is commensurate  \cite{Stone2001}, as is field-induced ordering \cite{Stone2006,Stone2007}. We therefore have to assume that incommensurability becomes favorable under pressure due to a continuously changing frustration ratios. In this context, the observed pressure-induced ordering in PHCC may be compared to a similar effect in the frustrated spin liquid Tb$_2$Ti$_2$O$_7$ \cite{Mirebeau2002}. The key difference is that in the latter compound the spin liquid state itself is due to geometric frustration \cite{Gardner1999}, and order seems to be induced by arbitrary low pressure. In PHCC, on the other hand, the distinctly gapped spin liquid state is due to spin dimerization, and persists all the way to the QCP at a non-zero $P_c$. 

Since pressure-induced ordering in PHCC is incommensurate,
the spin gap should soften at an incommensurate wave vector, similarly to what occurs at the {\it field}-induced incommensurate ordering transitions in the frustrated spin ladder Sul-Cu$_2$Cl$_4$ \cite{Garlea2009}. This would explain why neutron spectroscopy studies of Ref.~\cite{Hong2010} observed only a modest softening of the gap under pressure: they probed the spin gap at a commensurate wave vector.

The phase diagram of Fig.~\ref{Fig:phase} shows a typical multicritical point at $(P_1,T_1)$ , $P_1\sim 13.4$~kbar, $T_1\sim 2.0$~K. Here the paramagnetic phase meets the two ordered phases. The analysis described above indicates that the transition at $P_1$ is a IC transformation, so this is actually a potential Lifshitz point (LP) \cite{LL,Hornreich1980}. In spin systems, LPs are more commonly observed in applied magnetic field, as in the Dzyaloshinskii-Moriya helimagnet Ba$_2$CuGe$_2$O$_7$ \cite{Zheludev1997} and the quasi-1-dimensional Ising quantum magnet BaCo$_2$V$_2$O$_8$ \cite{Kimura2008}. Pressure-induced  Lifshitz points are known in ferroelectrics \cite{Kityk1999,Andersson2009} and metallic systems \cite{Zieba1991}, but are a more exotic occurrence in insulating quantum magnets. From our data on PHCC, it is not obvious if the transition below the LP (dashed line in Fig.~\ref{Fig:phase}) is a continuous one or not.  As to it's origin, it is probably driven by a continued change of the degree of frustration.

The observed QCP in PHCC is well within the accessible pressure ranges of many experimental techniques, including optical and neutron spectroscopies. This makes it an enormously important experimental prototype. First, even though the observed ordering is likely to be 3-dimensional, the quasi-two-dimensional structure and topology of magnetic interactions of PHCC may allow for a 2-dimensional quantum critical regime at elevated temperatures. The two-dimensional QCP case is unique in sense of the dynamics there can not be formulated in terms of quasiparticles. Instead, it is described in terms of ``quasi-normal'' modes with the language of conformal field theories, with surprising connections to black hole physics (\cite{Witczak2012} and references therein). To date, there is almost no experimental backing to the vast body of theoretical work in this area, and PHCC may provide the much needed opening. Second, the helimagnetic and possibly chiral 3-dimensionally ordered phase should couple to lattice degrees of freedom through reverse Dzyaloshinskii-Moriya interactions \cite{Sergienko2006,Mostovoy2006}. The result may be a novel transition from a spin liquid/ paraelectric state to a helimagnetic ferroelectric, and is of great fundamental interest. A field-induced version of such a quantum phase transition was recently observed in Sul-Cu$_2$Cl$_4$ \cite{Schrettle2013}. Additionally, in PHCC, further magnetoelectric features may be expected at the LP. Finally, the observed pressure effects turn PHCC into a promising model material for studying the effect of bond-disorder on the $SO(3)$ $z=1$ QCP. For a transition of this universality class, recent theoretical and numerical studies predicted very unusual features of excitation spectra and correlation functions \cite{Vojta2013}. At the same time, for PHCC, it has been demonstrated that bond randomness can be induced by chemical substitution on the halogen sites. The effect of this disorder on the {\it field-induced} transition in Br-doped PHCC \cite{Huevonen2012,Huevonen2012-2,Huevonen2013} has been discussed in terms of Bose Glass physics \cite{Zheludev2013,Huevonen2012}. Pressure experiments on this composition series will now allow to experimentally verify the predictions of Ref.~\cite{Vojta2013}, and to study the strongly inhomogeneous Griffiths regimes thought to occur near the transition.

This work was partially supported by the Swiss National Fund. It is a part of the PhD work of M. T. and of the MS thesis project of A. M.


\end{document}